\documentclass[reprint,
nofootinbib,
amsmath,
amssymb,
aps,
showpacs]{revtex4-2}

\usepackage[colorlinks,citecolor=blue,urlcolor=magenta]{hyperref}
\usepackage[T1]{fontenc}

\usepackage[inline,shortlabels]{enumitem}
\setlist[enumerate,1]{label=(\roman*)}

\newcommand{\Ncite}{154}
\newcommand{\diff}{\mathrm{d}}

\newcommand{\E}{\operatorname{E}}

\begin{document}


\title{`Ergodicity Economics' is Pseudoscience}
\author{Alexis Akira Toda}
\email{atoda@ucsd.edu}
\affiliation{Department of Economics, University of California San Diego}

\date{\today}

\begin{abstract}
In a series of papers \cite{Peters2011,PetersKlein2013,PetersGell-Mann2016,Peters2019}, Ole Peters and his collaborators claim that the `conceptual basis of mainstream economic theory' is `flawed' and that the approach they call `ergodicity economics' gives `reason to hope for a future economic science that
is more parsimonious, conceptually clearer and less subjective' \cite{Peters2019}. This paper argues that `ergodicity economics' is pseudoscience because it has not produced falsifiable implications and should be taken with skepticism.
\end{abstract}


\maketitle

\section{Introduction}

In a series of papers including ones that are published in prestigious physics journals, Ole Peters and his collaborators claim that the `conceptual basis of mainstream economic theory' is `flawed' and that the approach they call `ergodicity economics' gives `reason to hope for a future economic science that
is more parsimonious, conceptually clearer and less subjective' \cite{Peters2011,PetersKlein2013,PetersGell-Mann2016,Peters2019}. In \emph{The Logic of Scientific Discovery}, \citet[p.~18]{Popper1959} suggests that falsifiability should be taken as a criterion of demarcation between science and non-science. The Wikipedia article\footnote{\url{https://en.wikipedia.org/wiki/Pseudoscience}} defines pseudoscience as ``statements, beliefs, or practices that claim to be both scientific and factual but are incompatible with the scientific method''. The page lists the following indicators of possible pseudoscience:
\begin{enumerate*}
	\item\label{item:vague} use of vague, exaggerated or untestable claims,
	\item\label{item:evidence} improper collection of evidence,
	\item\label{item:open} lack of openness to testing by other experts,
	\item\label{item:progress} absence of progress,
	\item\label{item:person} personalization of issues, and
	\item\label{item:mislead} use of misleading language.
\end{enumerate*}
This paper argues that `ergodicity economics' advanced by Peters and collaborators is pseudoscience because
\begin{enumerate*}
    \item it has so far produced few testable implications and therefore is not falsifiable, and
    \item it meets many indicators of pseudoscience.
\end{enumerate*}

\section{`Ergodicity Economics'}

Henceforth we shall refer to `ergodicity economics' as EE. It is not easy to define what EE is because I am not an expert on this field (I am an expert on mathematical economics,\footnote{In case the reader questions my credentials, I hold an economics PhD (2013) from Yale University, I have published over 40 peer-reviewed research articles in economics and related fields, and I am a tenured associate professor at an economics department as well as a co-editor at \emph{Journal of Mathematical Economics} (\url{https://www.sciencedirect.com/journal/journal-of-mathematical-economics}).} which could be considered `mainstream economic theory'). Here, to avoid unnecessary debates, we shall focus on what is claimed in \citet{Peters2019}, which is published in \emph{Nature Physics} and is one of the most influential work of Peters by citation counts.\footnote{As of the time of writing, it is cited \Ncite~times.} It is not clear what kind of economic questions the author is trying to address (because the problem is never defined), but based on the discussion in the section titled ``A modern treatment asking the ergodicity question'', it seems that he is studying financial decisions under uncertainty, as he writes ``[r]eal-life financial decisions usually come
with a degree of uncertainty'' and ``[g]rowth rate optimization is now sometimes called {`ergodicity
economics'}'' (p.~1218). He also writes ``the geometric mean [\ldots] is well known among
gamblers as Kelly's criterion of 1956. Our modest contribution is to frame these observations as a question of ergodicity'', citing \cite{Kelly1956}. Based on this reading, I interpret that EE prescribes ``growth rate optimization'', that is, maximizing the geometric growth rate of an investment.\footnote{The idea of ``growth rate optimization'' dates back at least to \cite{Latane1959} and certainly before the treatment of optimal investment in a dynamic setting based on expected utility theory \cite{Merton1969,Samuelson1969}.}

\section{An optimal portfolio problem}

To contrast EE and the `mainstream economic theory' (expected utility theory, henceforth EUT\footnote{EUT was axiomatized by \citet{vonNeumannMorgenstern1944}; see their Section 1.3, and also \cite[Ch.~8]{Gilboa2009} for a textbook treatment. Although there are diverse views among economists regarding the theory of decision under uncertainty, most would agree that EUT is a useful benchmark. See \cite[Part III]{Gilboa2009} for models other than EUT.}) criticized in \cite{Peters2011,PetersKlein2013,PetersGell-Mann2016,Peters2019}, let us consider the following problem studied by \citet{Merton1969} and \citet{Peters2011}.

\subsection{Model}
There are two assets, risky (say a stock mutual fund) and safe (say a money market fund). An investor chooses a portfolio (proportion of money invested in each asset). What is the optimal portfolio? This problem seems to be fundamental for personal finance, and any theory of investment ought to provide some prescription. To make the problem concrete, suppose that the price $S_t$ of the risky asset evolves according to a geometric Brownian motion with drift $\mu>0$ and volatility $\sigma>0$, so the instantaneous net return\footnote{To be clear with technical terms, if an investment is multiplied by 1.1, we say that the gross return is 1.1. The net return is 0.1 or 10\%. The log return is $\log 1.1=0.0953$.} is
\begin{equation}
    \frac{\diff S_t}{S_t}=\mu\diff t+\sigma \diff B_t, \label{eq:GBM}
\end{equation}
where $B_t$ is a standard Brownian motion.\footnote{Some authors use $W_t$ or $W(t)$ to denote the Brownian motion (Wiener process) but I use the notation $B_t$ (which is equally common) because I will later use $W_t$ to denote wealth. As \citet{Peters2011} rightly points out in a footnote, ``[s]ome authors define the parameters of geometric Brownian motion differently [\ldots] notation must be carefully translated for comparisons''.} Thus, $\mu$ is the expected return. Suppose that the safe asset yields a risk-free return $0<r<\mu$. Suppose that at $t=0$, an investor with wealth normalized to 1 invests a fraction of wealth $\theta$ in the risky asset and the remaining fraction $1-\theta$ in the safe asset.\footnote{In the two-asset case, $\theta$ can be interpreted as the leverage. I refer to $\theta$ as ``portfolio'' because it has the natural generalization to multiple assets.} Although inessential, to make matters simple, suppose the portfolio $\theta$ is chosen once and for all and that it will be continuously rebalanced to keep this fraction. Then the investor's wealth at time $W_t$ evolves according to the geometric Brownian motion
\begin{align}
    \frac{\diff W_t}{W_t}&=\theta \frac{\diff S_t}{S_t}+(1-\theta)r\diff t \notag \\
    &=(r+(\mu-r)\theta)\diff t+\sigma\theta \diff B_t. \label{eq:Wt}
\end{align}
Finally, suppose the investor cares only about the terminal wealth $W_T$, where $T>0$ is the investment horizon. Applying It\^o's lemma to \eqref{eq:Wt}, log wealth evolves according to the Brownian motion
\begin{equation}
    \diff \log W_t=(r+(\mu-r)\theta-\sigma^2\theta^2/2)\diff t+\sigma\theta \diff B_t. \label{eq:logWt}
\end{equation}
Using the well-known property of the Brownian motion, the terminal wealth $W_T$ is lognormally distributed:
\begin{equation}
    \log W_T\sim N((r+(\mu-r)\theta-\sigma^2\theta^2/2)T,\sigma^2\theta^2T). \label{eq:lognormal}
\end{equation}

\subsection{Prescription of `ergodicity economics'}

EE prescribes to maximize the long run geometric average growth rate. Because we normalized the initial wealth to 1, the geometric average growth rate is simply $W_T^{1/T}$. Its logarithm is therefore normally distributed as
\begin{equation}
    \frac{1}{T}\log W_T\sim N(r+(\mu-r)\theta-\sigma^2\theta^2/2,\sigma^2\theta^2/T). \label{eq:geo_average}
\end{equation}
Note that the mean of the log geometric average growth rate (the arithmetic average log growth rate) \eqref{eq:geo_average} is constant at
\begin{equation}
    r+(\mu-r)\theta-\frac{1}{2}\sigma^2\theta^2, \label{eq:obj_EE}
\end{equation}
and the standard deviation $\sigma\theta/\sqrt{T}$ tends to 0 as the investment horizon $T$ tends to infinity. Therefore as $T\to\infty$, the log geometric average growth rate converges (in probability) to the mean \eqref{eq:obj_EE}. EE thus prescribes to maximize \eqref{eq:obj_EE}, and because the objective function is a concave quadratic function, the optimal portfolio is clearly $\theta=\frac{\mu-r}{\sigma^2}$. This solution was given by \citet[Equation (23)]{Peters2011}, though this type of reasoning goes back at least to \citet{Latane1959}.

\subsection{Prescription of `mainstream economic theory'}

In contrast, the `mainstream economic theory' approach---expected utility theory (EUT)---prescribes that the investor maximizes the expected utility $\E[u(W_T)]$,\footnote{This statement is an oversimplification of EUT. Economists have studied many other generalizations, for instance allowing for consumption in intermediate periods, non-independent and non-Gaussian returns, and portfolio adjustment over time.} where $u$ is called a von Neumann-Morgenstern utility function. The most commonly used utility function is the constant relative risk aversion (CRRA) utility function $u(x)=\frac{x^{1-\gamma}}{1-\gamma}$, where $0<\gamma\neq 1$ is called the relative risk aversion coefficient \citep{Pratt1964}. The case $\gamma=1$ corresponds to log utility $u(x)=\log x$. Using the property of the moment generating function of the normal distribution, for $\gamma\neq 1$ a straightforward calculation yields
\begin{align*}
    U(\theta)=& \E[u(W_T)]\\
    =&\frac{1}{1-\gamma}\exp\left(\left((1-\gamma)\left(r+(\mu-r)\theta-\frac{1}{2}\sigma^2\theta^2\right)\right.\right.\\
    &\left.\left.+\frac{1}{2}(1-\gamma)^2\sigma^2\theta^2\right)T\right).
\end{align*}
A monotonic transformation yields
\begin{align}
    V(\theta)&= \frac{1}{T(1-\gamma)}\log((1-\gamma)U(\theta))\notag \\
    &=r+(\mu-r)\theta-\frac{1}{2}\gamma\sigma^2\theta^2. \label{eq:obj_EUT}
\end{align}
If $\gamma=1$, a similar calculation yields the identical objective function \eqref{eq:obj_EUT}. Since $V(\theta)$ is a concave quadratic function of $\theta$, the maximum is achieved when $\theta=\frac{\mu-r}{\gamma\sigma^2}$. This solution was given by \citet[Equation (25)]{Merton1969}.\footnote{The notation is slightly different. Our $\mu,\gamma$ correspond to \citet{Merton1969}'s $\alpha,1-\gamma$ but this is inessential.}

\section{Is `ergodicity economics' testable?}

Obviously, the objective functions of EE \eqref{eq:obj_EE} and EUT \eqref{eq:obj_EUT} agree (for this particular optimal portfolio problem) when $\gamma=1$, so both theories make identical predictions in that particular case. Some economists have criticized the `growth optimal' approach (maximizing the long run geometric average) because it is consistent with expected utility maximization only in the special case of logarithmic utility \cite{Samuelson1971,MertonSamuelson1974}.\footnote{Interestingly, \citet{Peters2011} cites \citet{MertonSamuelson1974} out of context in a sentence discussing the ``advantage of [maximized time-average growth rate]'' (p.~1601). In fact, \cite{MertonSamuelson1974} is quite critical to ``growth rate optimization'', writing (p.~69) ``Recourse to the Law of Large Numbers, as applied to repeated multiplicative variates (cumulative sums of logarithms of portfolio value relatives), has independently tempted various writers, holding out to them the hope that one can replace an arbitrary utility function of terminal wealth with  all its intractability, by the function $U(W_T)=\log W_T$: maximizing the geometric mean or the expected log of outcomes, it was hoped, would provide an asymptotically exact criterion for rational action''.} That critique is valid \emph{conditional on accepting expected utility theory}, but is obviously unconvincing to those who rejects EUT in the first place. Others have criticized EE for misunderstanding EUT \cite{DoctorWakkerWang2020,FordKay2022}.

Here I criticize EE from a different point of view. If EE is a science, according to \citet{Popper1959} it must provide testable implications: the theory must be falsifiable. The optimal portfolio problem described above is one testable implication: in principle, we can gather data on the model parameters $\mu,r,\sigma$ and check whether investors choose the portfolio $\theta=\frac{\mu-r}{\sigma^2}$ prescribed by `growth rate optimization' of EE. Similarly, EUT is testable by perhaps conducting experiments in which we vary the model parameters $\mu,r,\sigma^2$ and check whether the chosen portfolio is proportional to $\frac{\mu-r}{\sigma^2}$ (only proportional, because the relative risk aversion coefficient is not directly observable).

What if we consider a different problem? Suppose an investor lives for three periods denoted by $t=0,1,2$, which could be interpreted as the young age, middle age, and retirement. The (non-financial) income of the investor at time $t$ is denoted by $Y_t$ (which could be a random variable). There could be two (risky and riskless) assets as before; the gross return of the risky asset between time $t-1$ and $t$ is denoted by $R_t$, and the gross risk-free rate is $R_f$. The investor could also choose consumption $C_t$. Then the problem is to choose the consumption-investment plan $(C_t,\theta_t)$ satisfying the dynamic budget constraints
\begin{equation*}
    W_{t+1}=(R_f+(R_{t+1}-R_f)\theta_t)(W_t-C_t+Y_t)
\end{equation*}
for $t=0,1$ given $W_0=0$. EUT (or I should just say `economic theory' because it does not need to be expected utility theory) has a prescription for this optimal consumption-portfolio problem. Any first year PhD student in economics will be able to formulate the problem mathematically. Given the assumption (or data) on the utility function, income process, and the return process, a sufficiently competent PhD student will be able to solve the model numerically and compare the model predictions to the data. If the 3-period problem is unrealistic, no problem; it could be 30 period or infinite horizon or whatever. My point is that given the assumptions we make, economic theory will provide a testable implication that is in principle falsifiable, and therefore it is a science.

I wonder what EE (growth rate optimization) has to say about this simple model of consumption and investment. \citet{PetersGell-Mann2016} note that ``[a] generalization beyond purely additive or multiplicative dynamics is possible, just as it is possible to define utility functions other than the linear or logarithmic function. This will be the subject of a future publication''. I am looking forward to seeing how EE would solve the above consumption-investment problem, though I am not optimistic as it has been 12 years since \cite{Peters2011} was published and we have not seen much progress (and the result in \cite{Peters2011} is itself a minor reformulation of \cite{Kelly1956,Latane1959}). Until EE provides testable implications, we should stay away from it as pseudoscience.

\section{Indicators of pseudoscience}

Let us return to Wikipedia's indicators of pseudoscience discussed in the introduction. Many of them apply to EE.

\paragraph*{\ref{item:vague} Use of vague, exaggerated or untestable claims}
\citet{PetersGell-Mann2016} make the grandiose claim ``[t]he concepts we have presented resolve the fundamental problem of decision theory, therefore game theory, and asset pricing'' without presenting any concrete application or  potentially testable implication.

\paragraph*{\ref{item:evidence} Improper collection of evidence}
The only experimental evidence \citet{Peters2019} cites is \citet{Meder2021}, which has also been criticized \citep{DoctorWakkerWang2020}. \citet{Peters2011} cites \citet{MertonSamuelson1974} out of context. \citet{Peters2011} states ``[i]n economics, a mistaken belief in ergodicity has produced widespread conceptual inconsistency'' citing the review article \cite{YakovenkoRosser2009}, though nowhere in that article such a statement can be found.

\paragraph*{\ref{item:open} Lack of openness to testing by other experts}
As of the time of writing, Peters lists 34 research papers listed on his webpage.\footnote{\url{https://sites.santafe.edu/~ole/publications.html}} Among them, 22 are published and 12 are unpublished working papers. Among the 22 published papers, only three are published in journals remotely related to economics (one in \emph{Decision Analysis}, one in \emph{Quantitative Finance}, and one in \emph{Journal of Income Distribution}). Thus, despite the label `ergodic economics', it is not reviewed by experts in economics.

\paragraph*{\ref{item:progress} Absence of progress}
Peters often cites only very old works such as Huygens, Bernoulli, \citet{Kelly1956}, and \citet{Samuelson1969}. EE does not seem to add anything new beyond \cite{Kelly1956}.
\paragraph*{\ref{item:person} Personalization of issues}
\citet{PetersGell-Mann2016} criticize with unnecessarily strong language: ``Menger did decision theory a crucial disservice by undoing Laplace's correction, adding further errors, and writing a persuasive but invalid paper on the subject that concluded incorrectly''.

\paragraph*{\ref{item:mislead} Use of misleading language}
The abstract of \citet{Peters2019} states ``Economics typically deals with systems far from equilibrium---specifically with models of growth. It may therefore come as a surprise to learn that the prevailing formulations of economic theory---expected utility theory and its descendants---make an indiscriminate assumption of ergodicity''. Here it is unclear what is meant by ``equilibrium'' (there are many general equilibrium models of growth in economics) and what ``indiscriminate assumption of ergodicity'' refers to.

\section{Concluding remarks}

Physicists and economists have a long history of interaction. Walras and Pareto, who have made fundamental contributions to economics, were also trained in physics and engineering. The first Yale economics PhD, Irving Fisher (1891), was a student of the physicist Willard Gibbs. More recently, since the late 1990s, there was a fruitful interaction between physicists and economists for studying the power law behavior \citep{LevySolomon1996,ManrubiaZanette1999,Gabaix1999,GabaixGopikrishnanPlerouStanley2006}. I myself have modest contributions published in physics journals \citep{Toda2011PRE,BeareToda2020PhysD} and acknowledge physicists' contributions to economics \citep{BeareToda2022ECMA}.

I appreciate the open-mindedness of physicists and welcome their perspectives on economic phenomena. However, I must express my concerns about a few researchers who lack proper training in the subject matter and attack economics based on ignorance and misunderstanding. Even more concerning is that some of them, like Peters, have used popular and social media to disseminate pseudoscientific information \cite{Bloomberg2020,Morningstar2020,EEblog,PetersTwitter}. We should all follow scientific methods, and I hope that scientific journals can maintain the integrity of peer review.

\bibliography{localbib}

\end{document}